\documentclass[prb,twocolumn,showpacs,floatfix,amsmath,amssymb,superscriptaddress,showkeys]{revtex4-1}

\usepackage{mathbbol}
\usepackage{graphicx}
\usepackage{epstopdf}
\usepackage{dcolumn}
\usepackage{bm}
\usepackage{array}
\usepackage{hyperref}
\begin{document}
	
\bibliographystyle{plain}
\title{Band structure reconstruction across nematic order in high quality FeSe single crystal as revealed by optical spectroscopy study}
\author{Haipeng Wang}
\affiliation{Beijing National Laboratory for Condensed Matter Physics, Institute of Physics, Chinese Academy of Sciences, Beijing 100190, China}
\author{Zirong Ye}
\affiliation{International Center for Quantum Materials, School of Physics, Peking University, Beijing 100871, China}
\author{Yan Zhang}
\affiliation{International Center for Quantum Materials, School of Physics, Peking University, Beijing 100871, China}
\affiliation{Collaborative Innovation Center of Quantum Matter, Beijing 100871, China}
\author{Nanlin Wang}
\email{nlwang@pku.edu.cn}
\affiliation{International Center for Quantum Materials, School of Physics, Peking University, Beijing 100871, China}
\affiliation{Collaborative Innovation Center of Quantum Matter, Beijing 100871, China}

\begin{abstract}
We perform an in-plane optical spectroscopy measurement on high quality FeSe single crystals grown by a vapor transport technique. Below the structural transition at $T_{\rm s}\sim$90 K, the reflectivity spectrum clearly shows a gradual suppression around 400 cm$^{-1}$ and the conductivity spectrum shows a peak at higher frequency. The energy scale of this gap-like feature is comparable to the width of the band splitting observed by ARPES. The low-frequency conductivity consists of two Drude components and the overall plasma frequency is smaller than that of the FeAs based compounds, suggesting a lower carrier density or stronger correlation effect. The plasma frequency becomes even smaller below $T_{\rm s}$ which agrees with the very small Fermi energy estimated by other experiments. Similar to iron pnictides, a clear temperature-induced spectral weight transfer is observed for FeSe, being indicative of strong correlation effect.
\end{abstract}
  \maketitle

\section{Introduction}
The PbO-type FeSe has the simplest structure in iron-based superconductors with a stack of edge-sharing $\mathrm{FeSe_4}$-tetrahedra layer by layer. The undoped FeSe exhibits superconducting transition at $T_{\rm c}\sim$9 K, but it can easily reach 37 K upon applying hydrostatic pressure \cite{FeSe-37K} and over 40 K by intercalating space layers  \cite{XFLu,HaiLin2016}. The $T_{\rm c}$ of the single layer FeSe film grown on $\mathrm{SrTiO_3}$ (STO) has been reported to be $\sim$60$-$70 K in angle-resolved photoemission spectroscopy (ARPES) experiments \cite{FeSe-STO-ARPES1,FeSe-STO-ARPES2,FeSe-STO-ARPES3,FeSe-STO-ARPES4} and Meissner effect measurements \cite{Zhang2015,Jia2015}. A recent in-situ transport measurement on monolayer of FeSe grown on SrTiO$_3$ even shows a sign of superconductivity over 100 K  \cite{FeSe-STO-109K}.

FeSe undergoes a structural transition at $T_{\rm s}$ = 90 K. However, different from the iron arsenic based systems, in which a tetragonal to orthorhombic structural transition usually precedes or coincides with stripe-type antiferromagnetic (AFM) order, there is no long-range magnetic order at any temperature in FeSe \cite{FeSe-Neutron,FeSe-NMR}. Experimental and theoretical studies suggested that the structural transition is caused by the electronic nematicity which breaks the $C_4$ rotational symmetry. It is widely believed that the mechanism of nematicity is tied to the pairing mechanism of high-$T_{\rm c}$ superconductivity in Fe-based superconductors. However, whether the nematicity has a spin or orbital origin remains controversial. On the other hand, FeSe superconductors also show several other peculiar physical properties. A recent study shows that FeSe has extremely small Fermi energy ($\varepsilon_{\rm F}$) and the superconducting energy gap ($\varDelta$) is comparable to the Fermi energy. This places FeSe superconductor in the Bose-Einstein condensation - Bardeen-Cooper-Schrieffer (BEC-BCS) crossover regime \cite{BEC-BCS}. An ARPES study \cite{Dirac-cone} revealed the existence of Dirac cone band dispersions in FeSe thin films, and an theoretical study \cite{JPHu} predicated a nematicity-driven topological phase transition in FeSe. In any sense, the nature of these fascinating phenomena and their interplay need to be understood comprehensively by using different experimental methods.

For a long time, the single crystal growth of FeSe with a pure ab-plane facet turned out to be difficult because of the narrow phase
formation range. The reported plate-like single crystals grown by flux method
usually contained secondary phase and quite often had a (101) surface \cite{Petrovic 101}, making it very difficult to study the intrinsic in-plane properties. As a result, pure FeSe  was far less studied than other Fe-based superconductors, particularly by spectroscopic techniques. Although some efforts were made in growing relatively thick ab-plane thin-films by pulsed laser deposition technique, which permitted measurement of some in-plane properties, e.g. by optical spectroscopy technique, the quality of the films grown by this technique was far from perfect. A substantial progress was made only after the vapor transport growth technique was used to grow FeSe crystals  \cite{AEB}. The high quality single crystal samples obtained by this technique enable researchers to investigate the various electronic properties by different techniques.

\begin{figure*}
	\includegraphics[width=17cm]{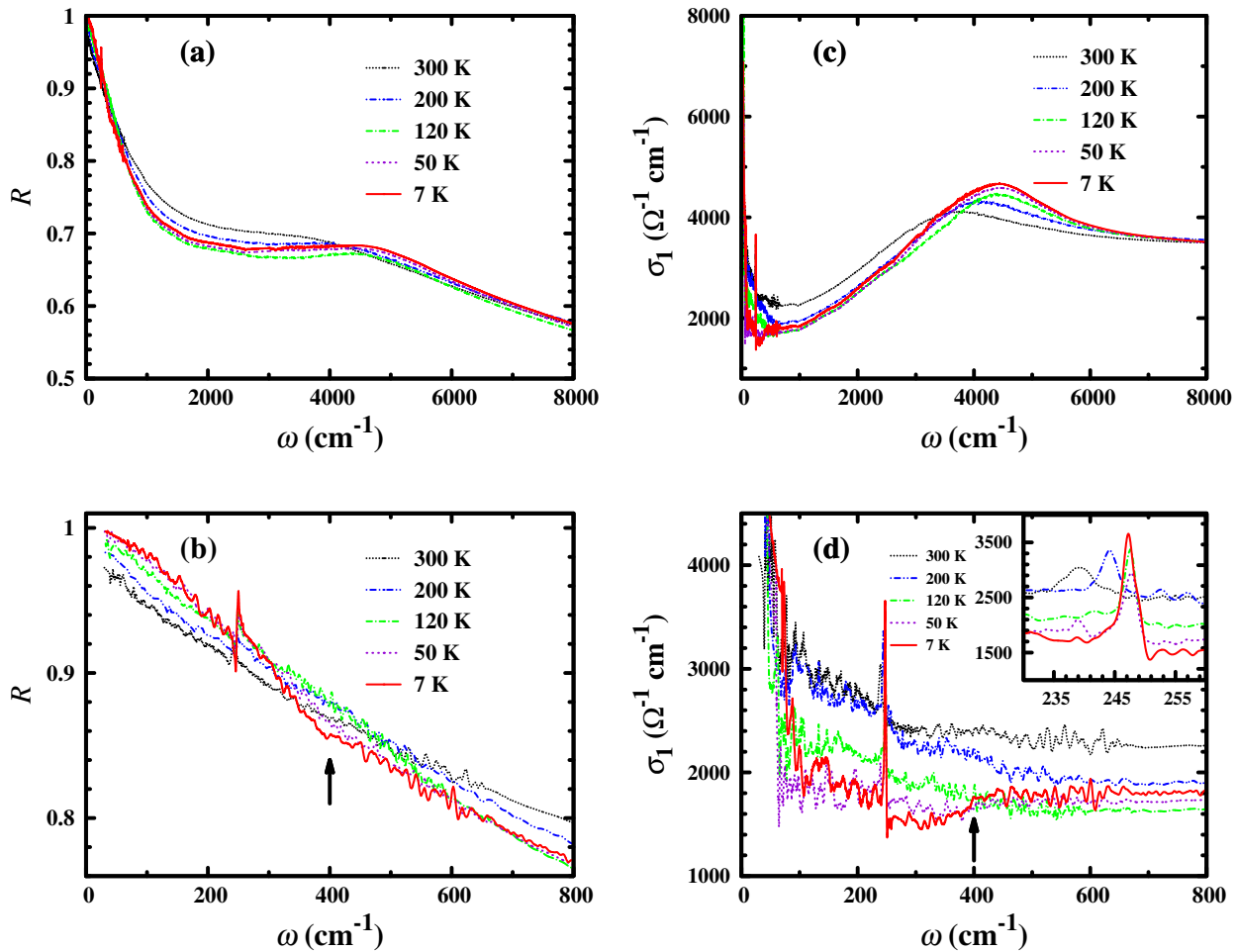}
	\caption{(Color online) \textbf{a} Optical reflectivity at five representative temperatures in the frequency range from 0 to 8000 cm$^{-1}$. \textbf{b} Optical reflectivity from 0 to 800 cm$^{-1}$. The arrow marks the position of the suppression of $R(\omega)$ below 120 K. \textbf{c} Optical conductivity $\sigma_1(\omega)$ at five representative temperatures from 0 to 8000 cm$^{-1}$. \textbf{d} Optical conductivity $\sigma_1(\omega)$ from 0 to 800 cm$^{-1}$. The arrow marks the position of the gap-like peak below 120 K. Inset of (\textbf{d}) shows the phonon peaks in the expanded scale   \label{R-sigma1}}
\end{figure*}
Optical spectroscopy is a powerful bulk technique to investigate charge dynamics and band structure of materials as it probes both free carriers and interband excitations. Our previous studies of FeSe thin film on STO \cite{FeSe-STO-Optic} has shown some interesting results but it is not enough to understand the complex physics in FeSe. No spectral change associated with the structural or nematic order was observed due to the relatively low quality of the thick films. So further optical studies of bulk FeSe are necessary. In this work, we report an optical spectroscopy study on high quality FeSe single crystal.

\section{Experimental}
The FeSe single crystals were grown using vapor transport method  \cite{AEB}. The optical reflectance measurements were performed on a combination of Bruker IFS 80 v/s and 113 v spectrometers in the frequency range from 30 to 26000 cm$^{-1}$. An \emph{in situ} gold and aluminum overcoating technique was used to get the reflectivity $R(\omega)$. The real part of conductivity $\sigma_1(\omega)$ was obtained by the Kramers-Kronig transformation of  $R(\omega)$ employing an extrapolation method with X-ray atomic scattering functions  \cite{KKXRO}. This new method of Kramers-Kronig transformation was proved to be more accurate and effective in analyzing the optical reflectivity.

\section{Results and Discussions}

The temperature-dependent in-plane optical reflectivity $R(\omega)$ and conductivity $\sigma_1(\omega)$ are shown in Fig. \ref{R-sigma1}. At low frequencies (the lower panels of Fig. \ref{R-sigma1}) a sharp phonon mode at about 240 cm$^{-1}$ can be seen clearly. This mode shifts slightly to higher frequency and becomes sharper with decreasing temperature. It can be seen more clearly in the inset of Fig. \ref{R-sigma1}d. This phonon mode is commonly seen in Fe-based superconductors and ascribed to the in-plane displacements of Fe-As(Se) atoms. The mode appears at a higher frequency than that observed at 187 cm$^{-1}$ for $\mathrm{Fe_{1.03}Te}$ and 204 cm$^{-1}$ for $\mathrm{FeTe_{0.55}Se_{0.45}}$ samples \cite{FeTe0.55Se0.45-Optic}. The effect is due to the reduced mass of the Se atom as compared with the Te atom. The clear observation of this in-plane phonon mode and its temperature-dependent effect indicates the good quality of the single crystal sample.

\begin{figure}[!htp]
	\includegraphics[width=8.5cm]{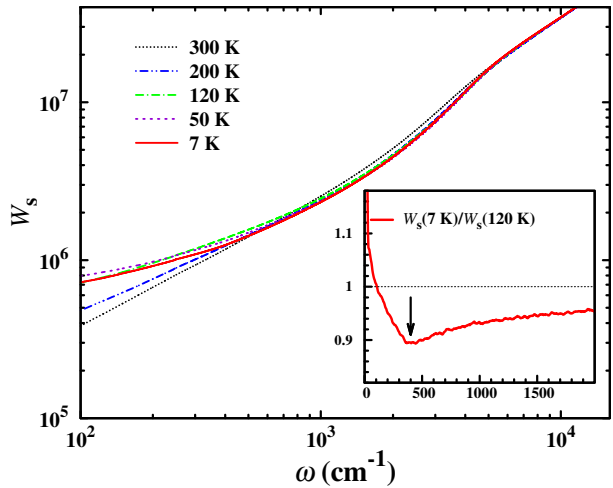}
	\caption{(Color online) Cutoff frequency dependent spectral weight at five different temperatures. Inset: the normalized spectral weight $W_{\rm s}$(7 K)/$W_{\rm s}$(120 K) up to 2000 cm$^{-1}$ \label{SW}}
\end{figure}

\begin{figure*}[!htp]
\includegraphics[width=16cm]{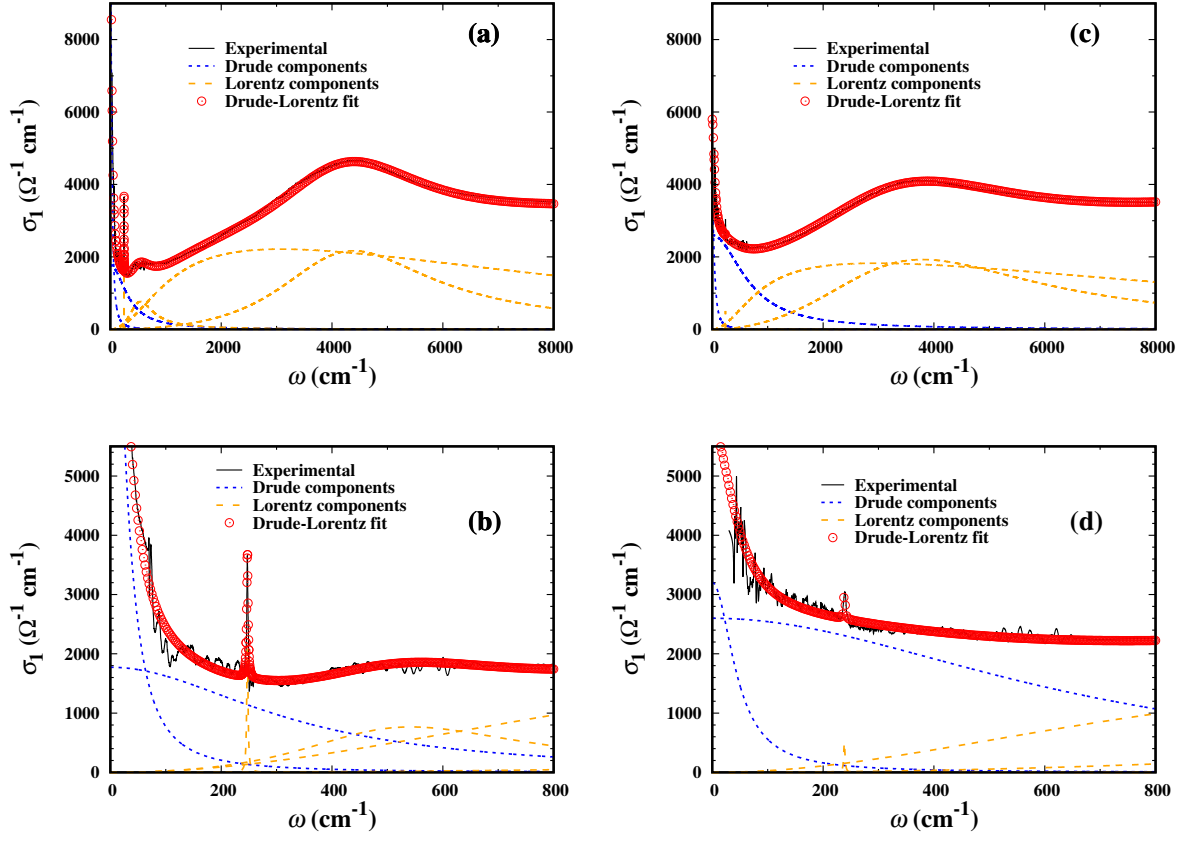}
	\caption{(Color online) The experimental data of $\sigma_1(\omega)$ together with the Drude-Lorentz fits at 7 K (\textbf{a} from 0 to 8000 cm$^{-1}$, \textbf{b} from 0 to 800 cm$^{-1}$) and 300 K (\textbf{c} from 0 to 8000 cm$^{-1}$, \textbf{d} from 0 to 800 cm$^{-1}$) \label{fit}}
\end{figure*}

Below the structural phase transition (in nematic phase) a clear temperature dependent suppression is seen in $R(\omega)$, leading to a dip-like feature near the frequency of 400 cm$^{-1}$ as indicated by an arrow in Fig. \ref{R-sigma1}b. Accordingly, $\sigma_1(\omega)$ shows a peak-like structure at somewhat higher frequency. This is a typical gap-like feature. Moreover, this feature can be seen from the spectral weight analysis as we shall present below (see inset of Fig. \ref{SW}). ARPES studies have revealed that there exists the splitting of ${d_{xz}}$ and ${d_{yz}}$ bands and possible additional separation of ${d_{xz/yz}}$ and ${d_{xy}}$ bands in nematic phase \cite{ARPES-Nakayama,ARPES-Chubukov,ARPES-HDing}. The energy scale of the band splitting at low temperature is about 50 meV at M point  \cite{ARPES-Nakayama,ARPES-Chubukov} and 30 meV at $\Gamma$ point  \cite{ARPES-HDing} which is comparable to the central energy of the Lorentz peak (540 cm$^{-1}$ or 67 meV) observed on $\sigma_1(\omega)$ (see Table \ref{tab:FeSe-Fit} and discussion below). The gap-like feature in our optical spectra thus should be associated with the band reconstruction below $T_{\rm s}$. Note that in the optical-conductivity measurements the gap value should be less than the location of the peak in $\sigma_1(\omega)$, thus the discrepancy between the results from ARPES and optics is small. It is also worth noting that a previous pump-probe experiment on FeSe film indicated an emergence of gaplike quasiparticles (associated with a energy gap $\varDelta\approx$ 36 meV) below 100 K  \cite{Pump-probe} though the extraction of an energy gap from this technique is indirect. This gap-like feature is similar as the energy gap formation in nematic phase observed by infrared experiment in parent compounds of iron pnictides  \cite{WZHu2008,Fe111-Optic,Fe1111-Optic} (including 122, 111, and 1111 systems). In these iron pnictides the nematicity is generally considered to be driven by magnetism \cite{Nematic}. However, unlike the iron pnictides, there is no stripe-type long-range antiferromagnetic order in FeSe at any temperature. Because the absence of long-range magnetic order and weak spin-fluctuation at high temperature around $T_{\rm s}$ observed by NMR \cite{FeSe-NMR} the orbital-driven model seem to be more plausible. This scenario was support by some ARPES experiments  \cite{ARPES-Nakayama} and some theoretical studies \cite{theory-OO}. However, a recent theoretical study based on J$_1$-J$_2$ model of spin-1 local moments with strongly frustrated exchange interactions indicated that its  quantum fluctuations can lead to a nematic quantum paramagnetic phase consistent with the observations in FeSe \cite{FWang}. A recent inelastic neutron scattering study \cite{FeSe-Neutron} revealed substantial commensurate stripe spin fluctuations in the tetragonal phase and there exists strong coupling between the stripe spin fluctuations, nematicity and superconductivity in FeSe. A polarized ultrafast spectroscopy study was also in favor of the magnetic origin of nematicity  \cite{Ultrafast}. Apparently, the nematicity which is responsible for the band reconstruction in FeSe have not yet well-understood. The gap-like feature observed in our optical results could give helpful information on this topic.

\begin{table*}[!htp]
	\caption{Parameters of Drude-Lorentz fit at low frequencies and the overall plasma frequencies at different temperatures }
	\label{tab:FeSe-Fit}
	\centering
	\begin{tabular}{p{1.2cm}<{\centering}   p{1.2cm}<{\centering}p{1.2cm}<{\centering}  p{1.2cm}<{\centering}p{1.2cm}<{\centering} p{1.2cm}<{\centering}p{1.2cm}<{\centering} p{1.2cm}<{\centering} p{2.5cm}<{\centering} p{2.9cm}<{\centering}}

    \hline
	&\multicolumn{2}{c}{Drude1} & \multicolumn{2}{c}{Drude2} & \multicolumn{3}{c}{Lorentz}& \multicolumn{2}{c}{$\omega_{\rm p}$}\\
		 $T$ & $\omega_{{\rm p}1}$ &  $\varGamma_1$   &$\omega_{{\rm p}2}$  &  $\varGamma_2$   & $\omega_{0}$  & $\omega_{\rm p}$  &$\varGamma_{}$  &{\footnotesize  ${(\omega_{{\rm p}1}^2+\omega_{{\rm p}2}^2)}^{1/2}$}     &{\footnotesize  $[8 \int_0^{\omega_c} \sigma(\omega) {\rm d}\omega]^{1/2}$}   \\
	(K)&(cm$^{-1}$)    &  (cm$^{-1}$)   & (cm$^{-1}$)    & (cm$^{-1}$)   & (cm$^{-1}$)   & (cm$^{-1}$)  & (cm$^{-1}$)  & (cm$^{-1}$)   & (cm$^{-1}$)    \\
	\hline
	300 & 10200 & 670 	& 3310  & 45   & -   & -	& -       & 10700  & 8900 \\
	200 & 10000 & 655 	& 3330  & 36  & -   & - 	& -       & 10500  & 8700 \\
	120 & 8800  & 640 	& 3500  & 36   & -   & -  & -         & 9400   & 8500 \\
	50 	& 6300 	& 409 	& 3600  & 34   & 510 & 4600 & 600     & 7300   & 6570  \\
	7 	& 5940 	& 330  	& 4160  & 30   & 540 & 4840 & 510     & 7200   & 6170 \\
	\hline
	\end{tabular}
\end{table*}

At low frequencies, Drude-like response exists in the conductivity spectrum (see Fig. \ref{R-sigma1}d). This behavior is also observed in Fe-pnictides such as $\mathrm{FeTe_{0.55}Se_{0.45}}$ and $\mathrm{BaFe_2As_2}$  \cite{WZHu2008,Fe111-Optic}. To estimate the Drude weight or plasma frequency of the FeSe and characterize the spectral change across the phase transition, we decompose the optical conductivity spectral into different components using a Drude-Lorentz analysis. The dielectric function has the form  \cite{Nakajima,WuDan,Homes,WZHu2008}
\begin{equation}
\epsilon(\omega)=\epsilon_{\infty}-\sum\limits_{k} \frac{\omega_{{\rm p},k}^2}{\omega^2+{\rm i}\omega/\tau_k} +\sum\limits_{j} \frac{\varOmega_{j}^2}{\omega_{j}^2-\omega^2-{\rm i}\omega/\tau_j},
\end{equation}
where $\epsilon_\infty $ is the dielectric constant at high energy, the middle and last term are the Drude and Lorentz components. The complex conductivity is $\sigma(\omega)=\sigma_1(\omega)+i\sigma_2(\omega)=-\omega[\epsilon(\omega)-\epsilon_\infty]/4\pi$. We use two Drude components (a sharp one and a broaden one) and several Lorentz components, including a Lorentz peak of phonon, to fit the optical conductivity. The conductivity spectra at different temperatures can be reproduced. The fitting results at two representative temperatures, the lowest one at 7 K and the room temperature, are shown in Fig. \ref{fit}. Detailed fitting parameters at low frequencies are listed in Table \ref{tab:FeSe-Fit}. Here, a Lorentz component at low frequency is added to character the gap-opening like feature of $\sigma_1(\omega)$ but not used above $T_{\rm s}$. Usually the energy of the gap can be estimated by the central frequency of this Lorentz model \cite{WZHu2008}. In this case the central frequency is about 540 cm$^{-1}$ (67 meV) at 7 K. In the two component approach, the overall plasma frequency $\omega_{\rm p}$ could be considered as being contributed from two different channels and be calculated as $\omega_{\rm p}={(\omega_{{\rm p}1}^2+\omega_{{\rm p}2}^2)}^{1/2}$. In the present case we obtain the $\omega_{\rm p}\approx$ 10000 cm$^{-1}$ above $T_{\rm s}$ and $\omega_{\rm p} \approx$ 7200 cm$^{-1}$ at low temperature as shown in Table  \ref{tab:FeSe-Fit}. Another method to estimate the overall plasma frequency is to calculate the low-$\omega$ spectral weight, $\omega_{\rm p}^2=8 \int_0^{\omega_c} \sigma(\omega) {\rm d}\omega $. The cut-off frequency $\omega_{\rm c}$ is chosen so as to make the integration cover all contribution from free carriers and exclude contribution from interband transition. Usually, the integral goes to a frequency where the conductivity shows a minimum. By choosing $\omega_{\rm c}$ (780 cm$^{-1}$ at 300 K, 750 cm$^{-1}$ at 200 K, 700 cm$^{-1}$ at 120 K, 360 cm$^{-1}$ at 50 K and 330 cm$^{-1}$ at 7 K) we recalculate the plasma frequencies and the results are shown in Table  \ref{tab:FeSe-Fit}.  All the values of $\omega_{\rm p}$ obtained by the second method are smaller than the former ones. Because the balance of Drude tail and Lorentz onset can not be accurately determined by the minimum of conductivity, the difference between two methods is not unreasonable.

The overall plasma frequency is related to the carrier density $n$ and effective mass $m^*$ by $\omega_{\rm p}^2 = {{ne^2} /({\epsilon_0 m^*}})$. Assuming that the effective mass of charge carriers does not change much with decreasing temperature, we get the ratio of $\omega_{{\rm p},7k}^2/\omega_{{\rm p},300k}^2$ (45  \% for first method and 48 \% for second method), suggesting more than half of the free carrier spectral weight is suppressed after the structural or nematic phase transition. So the carrier density of FeSe at low temperature becomes significantly small. We remark that the band reconstruction or gap-like suppression below nematic phase transition was not observed in our earlier study on FeSe film grown on SrTiO$_3$ substrate by pulsed laser deposition technique  \cite{FeSe-STO-Optic}, most likely due to the relatively poor quality of the film compared with present single crystal samples grown by vapor transport technique. The significantly small $\omega_{p}$ at low temperature is consistent with very small Fermi energy ($\varepsilon_{\rm F} \lesssim$ 10 meV) seen in recent ARPES and many other measurements \cite{SdH-prb-2014,SdH-epl-2015,ARPES-Chubukov,ARPES-HDing,ARPES-Nakayama}.  In fact, due to the very small $\varepsilon_{\rm F}$, it is suggested that FeSe is in the BEC-BCS crossover regime \cite{BEC-BCS,BEC-BCS-2} where $\varDelta/\varepsilon_{\rm F}\sim 1$.

As temperature decreases, from Fig. \ref{R-sigma1}c we can clearly see that the broad peak at $4000- 5000$ cm$^{-1}$ becomes narrow and shifts to higher energy. By defining the spectral weight as $W_{\rm s}=\int_0^{\omega_{\rm c}}\sigma_1(\omega) {\rm d}\omega$, the transfer of spectral weight can also be seen in Fig. \ref{SW}. $W_{\rm s}$ at low temperature between 2000 and 4000 cm$^{-1}$ is lower but recovers at higher frequency, \emph{i. e.}, the spectral weight is transferred to higher frequency region. This feature in the present high quality FeSe single crystal is more prominent than that observed in FeSe film on STO substrate \cite{FeSe-STO-Optic} and even other iron pnitctides \cite{SW-Trans,BaFeCoAs2-SW}. However, the relevant energy scale for FeSe is smaller than that for BaFe$_2$As$_2$ whose spectral weight was transferred to a region above 5000 cm$^{-1}$. Such temperature-induced spectral weight transfer was ascribed to the Hund's coupling effect between itinerant and localized Fe 3d electrons in different orbitals  \cite{SW-Trans}. FeSe as well as other iron-chalcogenide compounds have stronger electron correlation effect, leading to a larger local moment of 2$\mathrm{\mu_B}$  \cite{Fe-Moment} and higher band renormalization factor  \cite{Maletz2014}. The relatively smaller energy scale of temperature-induced spectral weight transfer reflects the complex interplay effect between Hund's correlation effect and kinetic energy of electrons in those iron-based superconductors.

\section{Summary}

In summary, we perform an in-plane optical spectroscopy measurement on high quality FeSe single crystals grown by a vapor transport technique. Below the structural or nematic phase transition at 90 K, the reflectivity spectrum clearly shows a gradual suppression around 400 cm$^{-1}$ and the conductivity spectrum shows a peak at higher frequency. The energy scale of this gap-like feature is comparable to the band splitting observed by ARPES and should be associated with the band reconstruction accompanied by the structural transition. The low-frequency conductivity consists of two Drude components and the overall plasma frequency is smaller than that of the FeAs based compounds. Further reduction of plasma frequency is observed below the structural phase transition temperature, being consistent with the very small Fermi energy observed in other measurements. Similar to iron pnictides, a temperature-induced spectral weight transfer is observed for FeSe, providing evidence for the strong correlation effect.

\begin{acknowledgments}
This work was supported by the National Natural Science Foundation of China (11120101003, 11327806), and the National Basic Research Program of China (2012CB821403).
\end{acknowledgments}

\textbf{Conflict of interest}

The authors declare that they have no conflict of interest.

\end{document}